# Guided mass spectrum labelling in atom probe tomography

*D. Haley[1,2], P. Choi[1], D. Raabe[1]*

1. Max-Planck-Institut für Eisenforschung, Max-Plack Straße 1, Düsseldorf, Germany

2. Department of Materials, University of Oxford, Parks Rd, Oxford, OX1 3PH, United Kingdom

Atom probe tomography (APT) is a valuable near-atomic scale imaging technique, which yields mass spectrographic data. Experimental correctness can often pivot on the identification of peaks within a dataset, this is a manual process where subjectivity and errors can arise. The limitations of manual procedures complicate APT experiments for the operator and furthermore are a barrier to technique standardisation. In this work we explore the capabilities of computer-guided ranging to aid identification and analysis of mass spectra.

We propose a fully robust algorithm for enumeration of the possible identities of detected peak positions, which assists labelling. Furthermore, a simple ranking scheme is developed to allow for evaluation of the likelihood of each possible identity being the likely assignment from the enumerated set. We demonstrate a simple, yet complete work-chain that allows for the conversion of mass-spectra to fully identified APT spectra, with the goal of minimising identification errors, and the inter-operator variance within APT experiments.

This work chain is compared to current procedures via experimental trials with different APT operators, to determine the relative effectiveness and precision of the two approaches. It is found that there is little loss of precision (and occasionally gain) when participants are given computer assistance. We find that in either case, inter-operator precision for ranging varies between 0 and 2 "significant figures" (2σ confidence in the first *n* digits of the reported value) when reporting compositions. Intra-operator precision is weakly tested and found to vary between 1 and 3 significant figures, depending upon species composition levels. Finally it is suggested that inconsistencies in inter-operator peak labelling may be the largest source of scatter when reporting composition data in APT.

## Introduction

Atom Probe Tomography (APT) is a powerful technique for obtaining 3D nanostructural data across a very small analysis volume, on the order of tens to hundreds of nanometres in each dimension [1][2]. APT is unique in combining atomic-scale chemical and spatial information, with data in the form of a "point cloud" with an associated mass-to-charge value for each point. The point cloud originates from the 3D reconstruction of discrete 2D detector events recorded during the experiment and can be on the order of $10^8$ events.

Labelling of the mass spectrum, much as in any spectrographic technique, is required to assign detected events to a particular atomic species. This step is nominally referred to in APT as "ranging", whereby each spectrum is assigned a separate "range file". Operators select a start and end for each range within the mass spectrum, as shown in Figure 1.

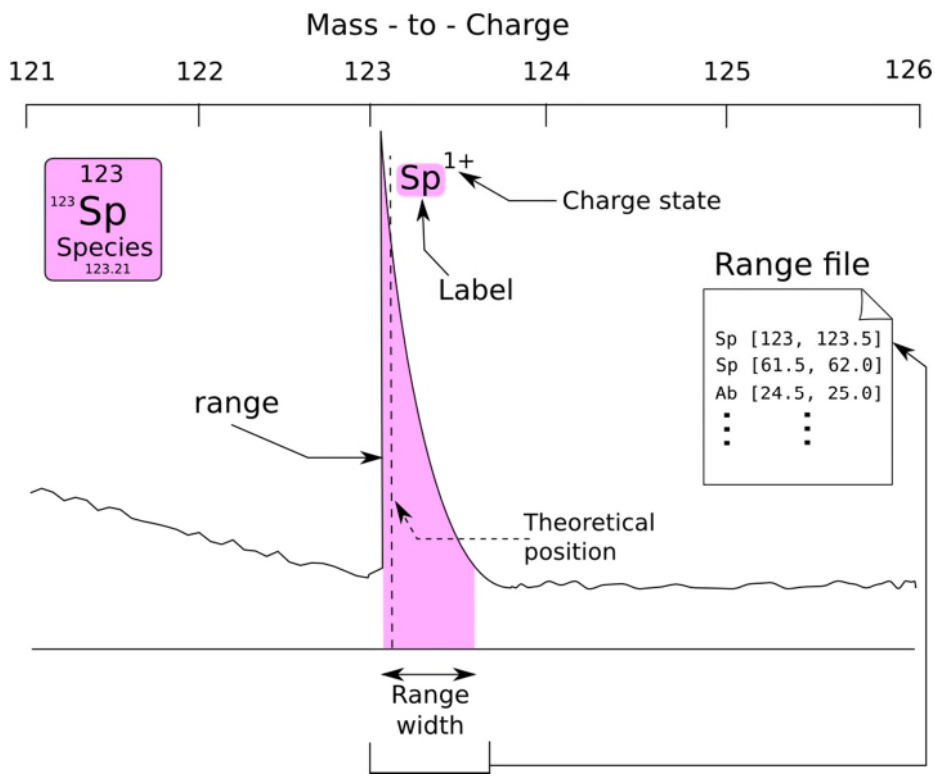

*Figure 1: Illustration of the ranging process: The operator labels a portion of the mass spectrum as a particular species. Selection is based upon isotope's (or combination thereof) theoretical mass and natural abundance.*

To estimate how frequently errors can arise in this process, two checks were undertaken on a corpus of 336 manually generated range files, each generated from different datasets. For each file, each range was checked to ensure that the ion label assigned to the range should be spanned; *i.e.* that the ion should have a peak in the assigned range, for any combination of isotopes. A second check, hereafter referred to as the "side-peak test" was undertaken to ascertain if one peak was assigned to a specific ion, then any theoretically larger peaks for that ion must also be assigned, however not necessarily to an ion of the same type (due to overlaps). Schematics indicating the tests are shown in Figure 2.

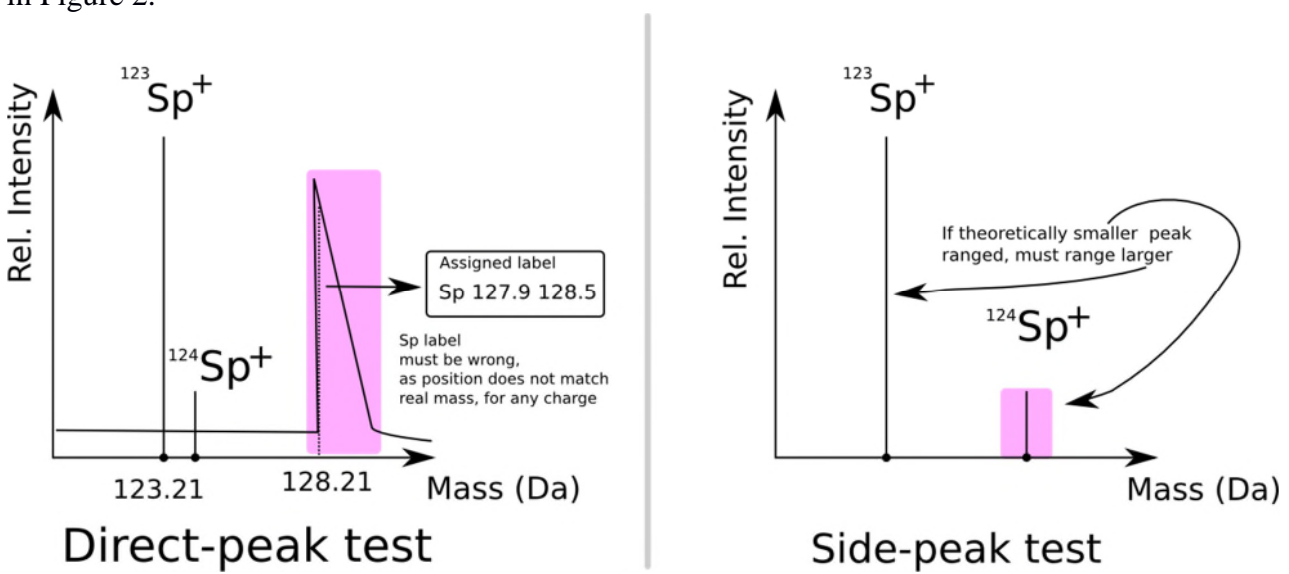

*Figure 2: Peak tests for direct mass existence, and for side-peak existence. Direct peak checks*

*label's theoretical mass lies in the specified range. Side-peak test checks that theoretically larger peaks must ranged if the smaller peak is ranged. Due to overlaps the identity of the larger peak's label may be different (i.e., something other than $Sp^+$).*

Each check was performed to within a mass window of ±0.1AMU, with charge states $1^+ \rightarrow 3^+$ (common charge states) being used to generate ion distributions. These checks have the capability to detect incorrect labellings, but are unable to determine the correctness of a given ranging.

From the above corpus, 87 of these files were found to possibly contain one or more errors according to these checks (~25% of the total). 168 ranges were marked as inconsistent (~2 per file), 12 of which were due to the side peak test, and the remainder due to the labelled species' isotopes not being located within this mass window (direct-peak test). Manual review of a random sample of 20 inconsistent files was conducted. 10 reports were confirmed, 9 were within ±0.4 AMU of the suggestion and were considered marginal, and one was due to 4+ charge state - thus a false positive.

Concerns about the appropriateness of ranging go beyond such simple checks, and have previously been tentatively discussed in APT literature [3], and are investigated here. It is the objective of this work to explore computational techniques to aid range selection, identification and validation, and to what extent this can be integrated into a single workflow. Such a workflow, if achievable, could reduce inaccuracies due to discrepancies between operators. Thus, robust methods in this direction can be seen as underpinning standardisation of data analysis of APT, and are the subject of this work.

## Proposed procedure for mass-to-charge peak labelling

During the analysis of a mass spectrum, operators are performing a *labelling* step, for each peak. Specifically, the task to be performed is that given a peak mass and the available atom types, the peak identity must be as *uniquely* determined insofar as possible.

This is usually based upon some context the analyst has, such as the expected elements. In the identification step, an analyst determines what possible elements can occur at a given mass-to-charge ratio. Common mass spectral techniques have highly complex mass spectra, as these are often used to investigate large-chain organic molecules, such as present in biological systems. These systems often utilise "fingerprint" database techniques to identify molecules based upon models of fragmentation behaviour of large chains [4][5]. Due to the relatively small chains of molecular species often present (usually of size 1) in atom probe mass spectra, such a fingerprinting method has limited applicability in an atom probe context. Here we examine highly robust methods that do not depend upon empirical databases.

Formally, the labelling problem can be expressed as the construction of a mapping from the set of possible species (or combinations thereof), *P*, to the set of true experimentally obtained (and hidden) species, *P'*. Note that as there may be multiple candidate solutions, there is often not a one-to-one mapping from *P*, to *P'*. In APT, the set *P* is constructed by building a set of species from elements' isotopes, extracted as a subset from the periodic table, then combining them to produce possible labels. It is desirable to minimise the size of *P*, as far as possible given any available information.

The set *P* can be constructed using the following rule: the masses from the isotopes, each having mass *m*, for each species present in *P*, must sum to an expected target mass-to-charge, *M*. The value of *M* being obtained from the mass spectra itself.

This problem is strongly related to a well explored mathematical problem, known as the *Knapsack Problem*, a particular problem that remains computationally complex [6]. The family of knapsack problems involves attempting to fill a fixed capacity container with a set of discrete weights of different size as completely as possible. In the context of APT the weights are the available isotopes, and the container is the target peak mass.

Optimal solutions to the knapsack problem, from a computational complexity perspective are at best solvable in linear time [7]. However such algorithms only provide one single solution to our posed problem, where for the problem here all solutions that fit the given tolerance must be elucidated. Thus we anticipate that it is improbable that there will be significant improvement, in terms of computational complexity, over a brute force solution.

From this perspective, a brute-force solution must be employed in order to elucidate the set *P*. To account for the difference of charge state, the knapsack search is repeated *N* times for each charge state *n*, with individual masses *m/n*, accumulating results into *P*. Pre-processing of the input can be used to preclude unnecessary searches (e.g. $m/n > M$), and drastically reduce search times by reducing the combinatorial space to be explored.

Practically, this means that a bounding set of molecular combinations must be made (maximum number of isotopes that can be combined in a single ion, *i*), and a maximum charge state, *n*, must be defined. Fortunately, for APT applications, the maximum values of *i*, and *n*, are usually sufficiently small to be computationally tractable. A small but simple modification to the knapsack problem must be made, as the true value of the mass-to-charge can only be approximated within some region of error, the brute-force solution must accept values to within some tolerance value, *ΔM*.

Thus, based upon a desired target mass *M*, a tolerance *ΔM*, the values of *i*, and *n*, and finally a list of elements, *E*, the set *P* can be built by brute force. For reasonable ranges of *i*, and *n*, ($i <=6$ and $n <=5$) and input set size *E* (~30 isotopes), a complete exploration of all possible combinations can be built, on timescales on or less than the order of seconds. It is important to note *ΔM* is not the peak width observed in the experiment, but is rather the uncertainty in the estimation of a given peak's mass centroid, and the alignment of the spectrum itself. For an appropriately calibrated mass spectrum, *ΔM* is usually far smaller than the width of a given peak.

Whilst it is possible to generate the set *P*, this in itself is insufficient for analysis, as the set *P* can be exceedingly large. Thus the method must further exclude entries that, whilst satisfying the condition $M - \Delta M <= \Sigma m <= M + \Delta M$, violate other physical principles of an atom probe experiment.

The reduction in set size is most effectively achieved by reducing the possible input set of isotopes, and by limiting the size of *i*, *n* and *ΔM*. However, such an approach can still yield an excessive size of *P*, for realistic input values, or the input parameters may be so narrowed as to exclude real solutions. Practically the size of *P*, prior to any subsequent reduction can often be on the order of dozens of possible solutions, or even more.

There exist several possible avenues for further reducing the size of *P,* given sufficient information. Knowing the stability of specific ions, or the ability to predict charge state, or at a minimum exclude impossible charge states would allow *P* to be reduced. However, stability and charge state questions are highly challenging to generalise, and, to the author's knowledge, no such model exists that is proven to be applicable in all cases – methodologies such as Kingham curves and evaporation fields often fail to predict charge state, e.g. under laser illumination [8] or in the presence of oxides [9].

One method that can be employed is the examination of side-peaks (Figure 2). If a possible solution species is thought to be a solution, the solution species will naturally have a set of peaks *K,* the distribution arising from the set of isotopic abundance values, $a_j$. This isotopic mass distribution can be approximated (assuming random mixing) by Equation 1, which states that the un-normalised probability of a peak, P(K'), of several differing isotopes, K'$_e$, is simply the product of the natural abundances for that peak. *e* and *f* index isotope combination and abundance value respectively.

$$P(K'_e) = \prod a_f ; K'_i = \{a_1, a_2, \ldots\}$$   Equation 1

If a peak K'$_e$ is observed, all peaks of larger intensity, i.e. (P(K'$_e$)/P(K'$_{e'}$)) >= 1, should exist. If the algorithm is supplied with a list of "observed" peaks from the real mass spectrum (without intensity), then the theoretically larger intensity peaks must be labelled for the solution to be valid. Due to overlaps, there is no guarantee that these peaks must have the same identity as P(K'$_{e'}$), only that they must be present.

Practically, for examined datasets, this test reduces the number of matches drastically. As an example, searching for the element list, {Fe,H,Mn,Si,Cu,Cr,C,B,P,Al,Ti,O,Na} at mass 24, with ±0.1 amu uncertainty, at *i*=3, *n*=3, initially yields ‖P‖ = 65 solutions (peaks from easy table, Table 1). After reduction of *P* via side-peak removal, only 11 possible solutions remained. However, additional information must now be robustly injected into the analysis, *i.e.*, the set of all observed peaks, *K*.

Lastly, even with the reduction of size *P*, then the labelling may remain ambiguous. Indeed, in the case of overlaps, it is theoretically impossible to reduce ‖P‖ to 1 by definition. In the absence of a more thorough method of reduction of *P,* heuristic methods can be employed to assign a "score" to each possible member of *P*, to provide a ranked list of possible candidates. Such a ranked list allows for more rapid rejection of unlikely combinations, such as mass 2 being suggested as $(^2H)_2^{2+}$, rather than the more probable, $H_2^+$

For the purposes of this work, a highly simplistic weighting scheme is utilised to roughly separate highly unlikely from possible elements. To do this, an assumed bulk composition multiplied by the natural abundance is used to assign a relative weight to the occurrence of each isotope. The product of the isotopic scores is the score for the final molecular ion. As an example, for an Fe-Mn alloy with 20at% Mn, the score for an $^{56}Fe^{54}Fe$ molecular ion would be (0.8*0.917)(0.8*0.058) = 0.034, assuming no contaminant species. For elements typically not present in the bulk, but present in the analysis as contaminants (eg H and O), a weighting factor must be given, based upon the propensity of the material (in the case of H) to be present in the APT dataset – unfortunately, estimations for this can be quite arbitrary. However, as the quality of the ranking is only heuristic and not exact in nature, inaccuracies below order-of-magnitude levels often do not change the relative ranking of the elements of *P*.

Whilst it is now possible to generate a suggestion set *P*, and roughly rank *P* using compositional data, additional data regarding peak positions is required for the reduction step. This places an additional burden on operators – however this too can be partially automated. Unlike the peak identification stage, APT mass spectral peak detection fundamentals are not too dissimilar to other mass spectral methods. It has long been considered that peak detection methods can be effective in correctly extracting peaks from a mass spectral signal [10], specifically in the related mass spectral imaging technique of MALDI-TOF.

Such peak extraction methods compute the total peak area without a-priori peak shape assumptions. Indeed, considerable work has been conducted in the area, with comprehensive reviews of the relative strengths and weaknesses of these automated approaches [11]. In techniques such as MALDI-TOF, mass spectra are highly complex [12], and peaks can be present in high mass regions, such as ~10,000 Dalton [13].

The MALDI-TOF tool "MaldiQUANT" was selected for the use in this work for the purposes of peak and background extraction [14], as it has been extensively developed. Comparative reviews for various signal processing techniques (such as wavelets [15], MEND [16]), are further discussed in detail elsewhere [17]. The optimisation of the peak extraction and identification steps for the context of APT are outside the scope of this work – peak detection here is used only to demonstrate the complete processing chain.

Similar to the method of Andreev [16], MaldiQuant was used to process time-domain signals, rather than the m/z domain, due to the non-linearity of the transform between the two domains, which results in artificially altered peak and background shapes. The output from MaldiQUANT which is relevant to this work is a set of detected peaks, and a background spectrum.

For this work, the method used was the wavelet "TopHat" mode [14], with a fixed Half-Width of 0.3 $AMU^{1/2}$. The cutoff amplitude for thresholding the wavelet-processed signal was set by first manually ranging, then reducing the cutoff until the same number of ranges (within a small tolerance ~2 peaks) was identified by the automated detector as for the manual ranging. In a full implementation, pre-calibrated thresholds can be used. Automatic identification was performed on the peak positions, and the identity was assigned as the highest ranked species from the set of suggestions for each peak.

## Method efficacy

To determine the efficacy of the method, both intra- and inter-operator variance is examined in the assisted and unassisted cases. Hereafter we refer to the generalised concept of a person operating the software as "operators". For specific people that undertook the study, we refer to them as "participants". Thus we estimate inter and intra-operator precision via a measurable inter and intra-participant error.

Three datasets were selected for comparison, with the datasets names were chosen based upon their perceived complexity for automatic identification. An "easy", "medium" and "hard" case given in Table 1 (PF: Pulse Fraction, TD: target detection rate). Each dataset was acquired from the same LEAP 3000X HR. All datasets masses were randomly shuffled prior to analysis, to eliminate spatial markers that some participants might use to aid their analysis.

The datasets were selected to be different to one another, to demonstrate that the method works in a more generalised fashion than for a single dataset. Furthermore we attempt to examine both fully metallic and ceramic materials (i.e. AlN). Hence we chose three datasets, with differing levels of apparent complexity for the complete algorithm (peak identification and labelling). As a further constraint to our selection, we limited our choice of dataset to ones which can be analysed by participants within a reasonable time frame, in order to ensure that analysis would be conducted in a complete fashion by all participants.

The relative difficulty in the datasets was, based upon tests conducted with the available datasets, due to the difficulty of peak identification. During the peak identification step, false positives and false negatives, particularly in peak tails can occur. As this work focusses on demonstrating a complete analysis chain and any potential benefits that may arise from this, optimisation of the peak-identification step will be addressed in a later work.

*Table 1: Selected materials and run conditions for operator precision case-study*

| Case | Material | Run Conditions |
| --- | --- | --- |
| Easy | Fe-Mn-Al-C steel | Voltage, 0.15 PF, 200 kHz, 70K, 0.5% TD |
| Medium | Fe-Ni-Mo-Ti Maraging steel | Laser, 0.4nJ, 100kHz, 80K, 0.5% TD |
| Hard | AlN/CrN Multilayer | Laser, 0.6nJ, 250kHz, 60K, 0.5% TD |

The easy case contains no major ambiguities in ranging, but does have an Al/Fe overlap at position 27 Da. Decomposition showed a 1.1:0.75 ratio of Fe:Al in this position (~60% total Fe). Manual and automated ranging attempts (intra-participant) ranged this as Fe. For the medium case, the signal-to-noise level was much lower than the easy case, and many molecular species were present (eg $MoO_x$) though no major overlaps were present. For the hard case, many overlaps were present, and participants were free to select the dominant range bounds. For the reported results, decomposition of overlaps is not considered when computing compositions – only the participant label is used.

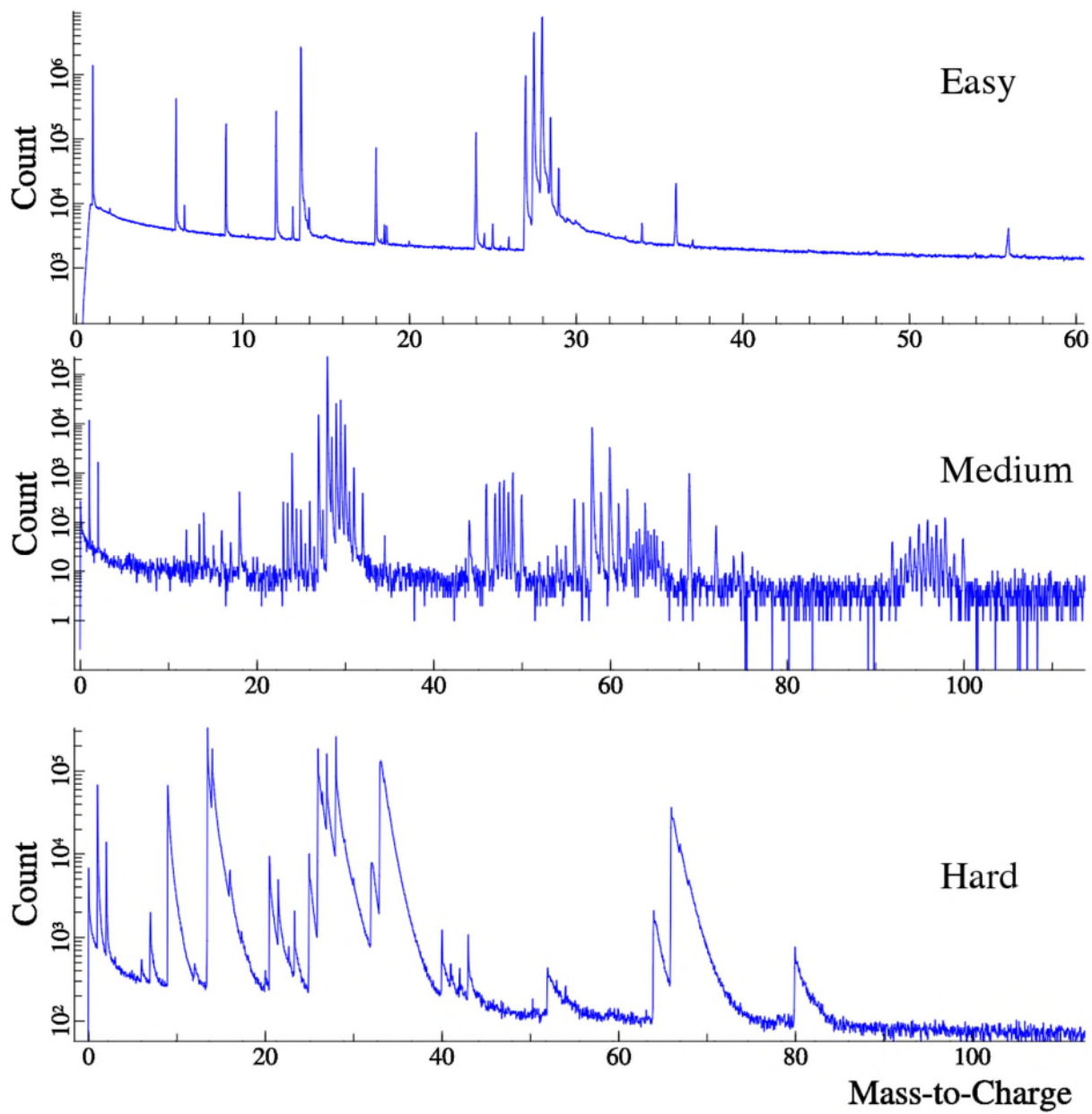

*Figure 3: Mass spectra for the selected cases. In the easy case, each peak is well separated, and has good signal. The medium and hard cases have lower signal and poor separation of peaks respectively*

**Intra-operator variance**

Using the "easy" dataset from Table 1, the author undertook six separate ranging attempts, each attempt separated by at least ten minutes. This was used to generate a composition for the overall dataset, after excluding the first "Test" ranging. The significant figure precision, which estimates the number of "accurate" leading digits in the reported composition value, is shown in Figure 4 -- computed as $D=-\log_{10}(2\sigma/C_i)$ of the composition ($\sigma$ – std. deviation). As an example, if the composition was reported to be 12.345%, and D =2, then the composition is 12± 0.5% assuming a Gaussian error distribution and a 2σ error threshold. As a further example, if the composition is 10±5%, then D = 0, as the leading digit is "inaccurate". Intra-participant composition was computed using IVAS 3.6.6, including background subtraction. Subsequently variation in measurement is purely single-participant precision, as the mass spectrum has not been altered.

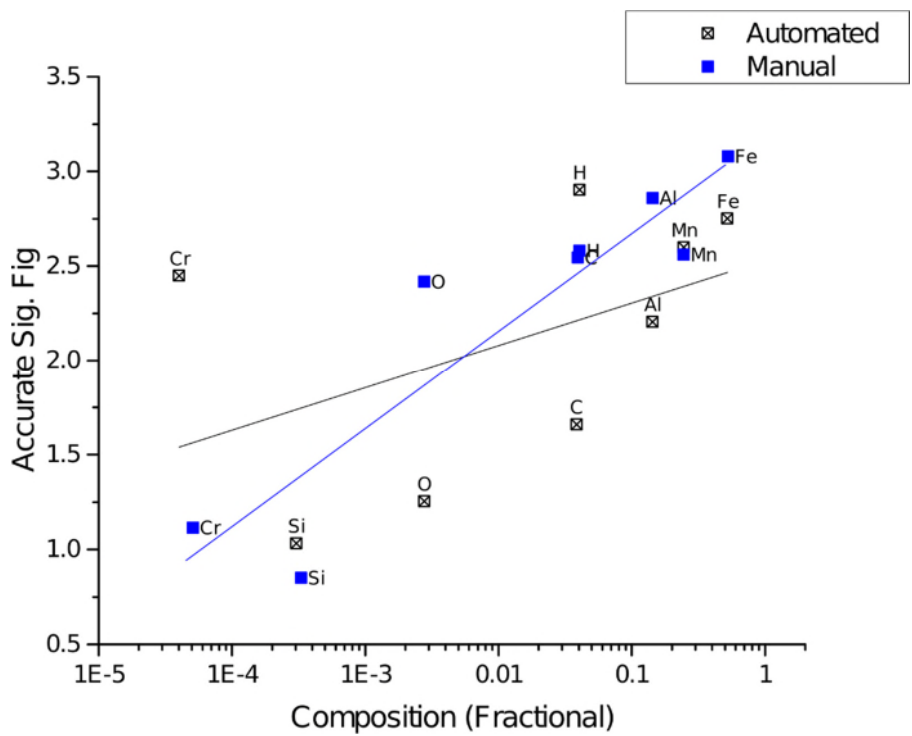

*Figure 4: Number of accurate significant figures (leading digits) in reported composition over repeated ranging attempts by the author (five trials) as a function of total composition, for an Fe-Mn-Al-C alloy (same dataset) for both computationally aided and unaided cases. Line fits are provided as a guide to the eye.*

As can be seen from Figure 4, there is a participant variance between each attempt. In the manual case the trend of single-participant precision is such that more dilute species have a lower precision, roughly linearly on a log-log scaling. H when using an automatically suggested rangefile as the starting point for ranging, the reported error has been reduced for H, Si, Mn and Cr; with Cr and H markedly reduced as during ranging, and this is ascribed to peaks being considered to be sufficiently well-labelled that they were minimally altered by the participant, thus providing a smaller quantity of scatter.

The mean time required for ranging was 8.0 minutes in the manual case, and 7.0 minutes in the assisted case, implying a slightly faster evaluation of datasets even without a customised user interface.

It is suggested that inter-operator variance will be higher when performing manual ranging. The results from Figures 4 show that there is little alteration in the intra-participant variance between the two cases, unless the participant chooses not to alter the peak at all. As a highly rough approximation for the achievable intra-operator variance E(x), during the ranging procedure the following equation is suggested based upon the logarithmic guide for the unaided case, where x is in atomic fraction, and E(x) in significant figures of precision at 2σ confidence. This equation is valid for this dataset, and is unclear how applicable this is to dissimilar systems.

$E(x) = 0.52 \log_{10}(x) + 3.2,$ Equation 1

As an example, using this equation, at 100 ppm, intra-operator precision is reduced to 1.2 significant figures. If this procedure were fully automated to an operator-acceptable level, the inter-operator error source could, in theory, be eliminated.

## Inter-operator variance

To collect data, participants were instructed to perform ranging using the IVAS program (3.6.4/3.6.6). Participants were assigned all three datasets, and were tasked with ranging the datasets twice – once starting from an empty set of ranges, and once using a pre-made rangefile, generated using the automated method. All participants were familiar with the IVAS program and with ranging atom probe data, however most participants were unfamiliar with the materials presented, and participants were not informed about the origin of the dataset, beyond its nominal composition, a list of likely elements and laser/voltage operation. Finally, all participants were from the same laboratory, and thus inter-laboratory estimations could be higher.

To minimise concerns about the effect of a-priori knowledge on the comparisons, the sequence (automated then manual, or vice-versa) was assigned randomly per participant, per dataset. For the purposes of the test, participants were explicitly instructed to range H and H-containing peaks as such, and to label multi-isotopes using their multiple form. No direction was given as to *how* peaks should be ranged. The authors were not included as participants in the study, and all participants had experience with APT in their own work, and self-declared as proficient with using the IVAS software package.

Thus up to six rangefiles were collected per participant, prior to filtering. A summary of data collection, after filtering, is given in Table 2. In the manual case, participants started with an empty ranging. In the assisted case, participants were supplied with an automatically generated ranging, and a list of possible labellings, as generated using the previously outlined method.

*Table 2: Summary of number of range-files obtained for each dataset, after filtering*

|          | Easy | Medium | Hard |
|----------|------|--------|------|
| Assisted | 11   | 9      | 9    |
| Manual   | 11   | 9      | 9    |

Resultant range files were removed from the analysis under several conditions:

- Non-unique files were wholly excluded from the analysis. One participant for one dataset submitted the same file for assisted and unassisted cases. – (2 files)
- Files which contained > 30% of peaks labelled as unknown – (2 files)
- Files that failed mass checks (all peaks should fit in the assigned mass to +-0.1 amu), e.g., if labelling a peak as H, it should lie within [0.9,1.1] or [1.9,2.1] amu (I.e. $^1$H or $^2$H) (1 file)
- Files whose total ranged count exceeded the mean total ranged count by ±2 standard deviations (3 files - caused by not labelling large peaks)

This rejection was done to minimise deviation from sources other than direct ranging errors (e.g. by not labelling H, and thus biasing composition comparisons between participants).

For background subtraction, a simple but largely effective subtraction model was used, where a horizontal line in TOF space is fitted [18] and converted to mass-to-charge space. The TOF data within a small window was tested for being sampled from a Gaussian distribution and could not be rejected as non-Gaussian ($P \ll 0.005$) according to the Anderson-Darling test [19]. The windows used for fitting were 1.3-1.9 Da for the easy case and 1.5-1.9 Da for the medium and hard cases. Compositions are computed by subtracting the area under the parameterised fit function, given by Equation 2, where $y_{fit}$ is the fitted TOF value, $x$ is the mass space, and $a$ and $b$ are the upper and lower mass bounds for background computation.

$$y_{\text{int}} = y_{fit} \left[ x^{\frac{1}{2}} \right]_a^b,  \qquad \text{Equation 2.}$$

**Result Comparison**

Total numbers of ions present in the dataset, and the number accurate significant figures are given in Table 3. Excepting the "Hard" dataset, no major alterations in precision of total count is seen between manual and assisted cases.

*Table 3: Summary of ranged number of ions and associated error for inter-operator comparisons. The total count error measures the agreement in the ranged number of ions.*

|  | Easy/Man. | Easy/Assist | Med./Man. | Med./Assist | Hard/Man. | Hard/Assist |
|---|---|---|---|---|---|---|
| Mean Count | $3.9 \times 10^7$ | $3.9 \times 10^7$ | $4.8 \times 10^5$ | $4.8 \times 10^5$ | $4.3 \times 10^6$ | $5.0 \times 10^6$ |
| Std. Dev. | $1.2 \times 10^6$ | $1.2 \times 10^6$ | $1.3 \times 10^4$ | $1.4 \times 10^4$ | $1.9 \times 10^6$ | $1.9 \times 10^6$ |
| Acc. SFs. in total count | 1.23 | 1.20 | 1.26 | 1.23 | 0.04 | 0.11 |

The spread of results obtained from participants is shown in Figures 5, 6 and 7 for each dataset. Only species which more than 5 participants identified, and are present in both the manual and automated datasets are shown.

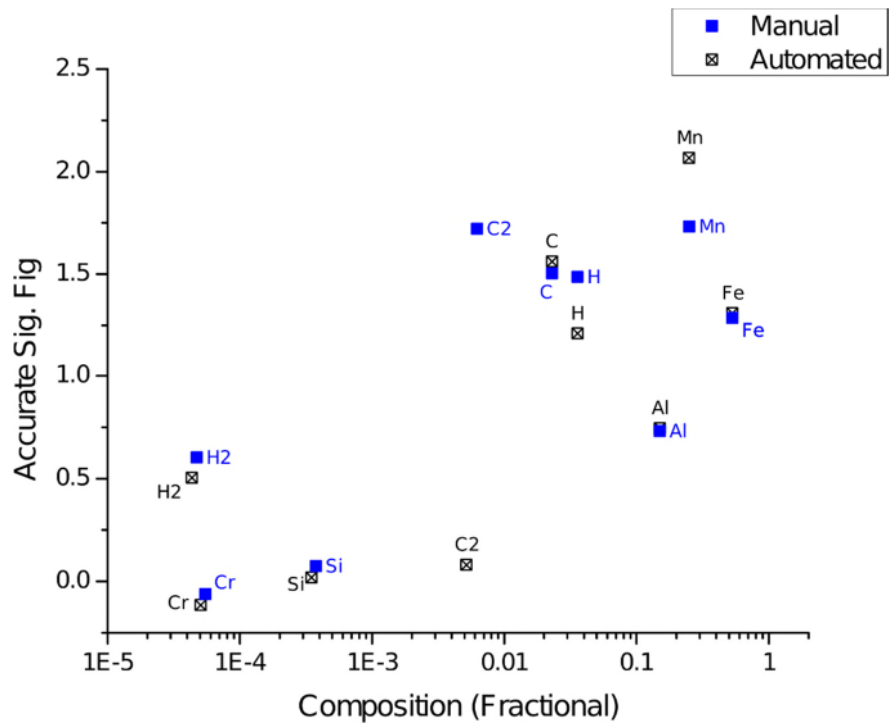

*Figure 5: Results from the Easy dataset comparison, using auto-assisted and manual ranging methods. Major changes in precision are only visible in C2 and Mn. C2 change is due to FeO/C2 misranging (as suggested by automated solution). Ranging of Mn has improved by using the automated method due to improved consistency in labelling.*

Figure 5 shows the "Easy" comparison. Aside from a slight increase in Mn inter-operator agreement which is due to more consistent peak labelling, additionally there is a marked decrease in $C_2$ precision. As in the automated algorithm, FeO was suggested as "more-likely" (23.976 vs 24.00), despite the absence of smaller side-peaks (smaller side-peaks are not accounted for in the current algorithm). In the automated case, several participants did not alter this default labelling leading to a bifurcation in the reported composition of $C_2$.

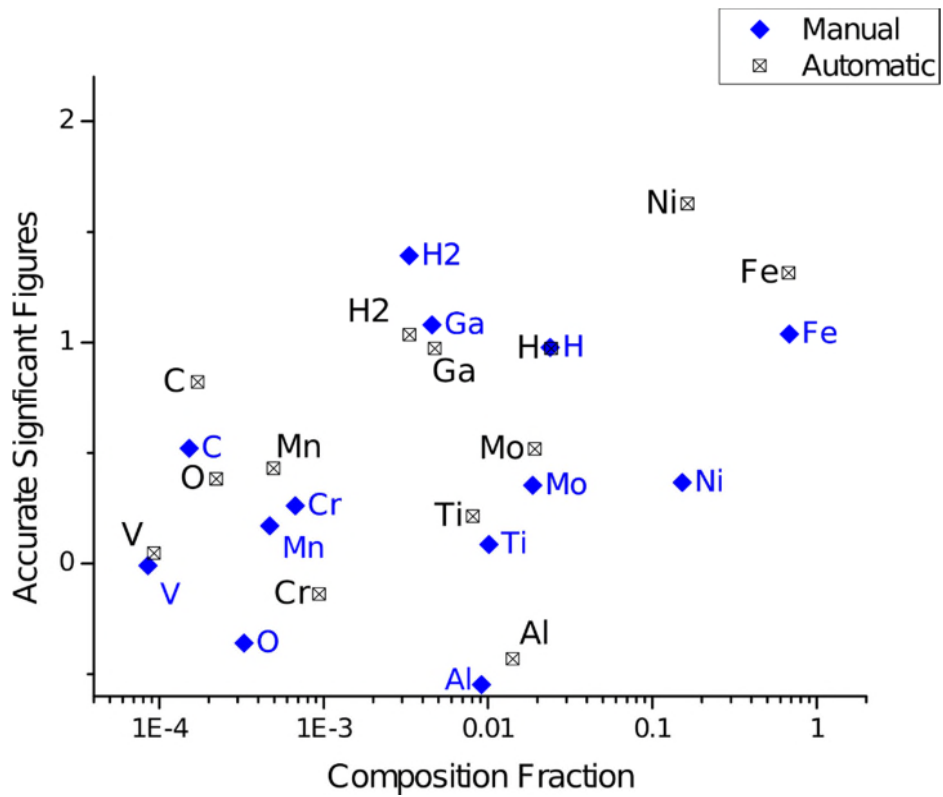

*Figure 6: Distribution of significant figures for the Medium case - Most species (excepting $H_2$ and Cr) show an improvement in inter-operator agreement, particularly Ni – where in the manual case, disagreement between Ni and Fe ranging was present.*

Figure 6 shows the "medium" dataset, for which two features are visible. Firstly, there are a large number of alloying elements within the host matrix, complicating peak labelling. Secondly, the total number of events in the dataset is relatively small. Combined with the first point, this provides a large number of low-signal peaks within the dataset, and thus low precision for many species. In this dataset, modest improvements are found for most species, with the exception of $H_2$ and Cr.

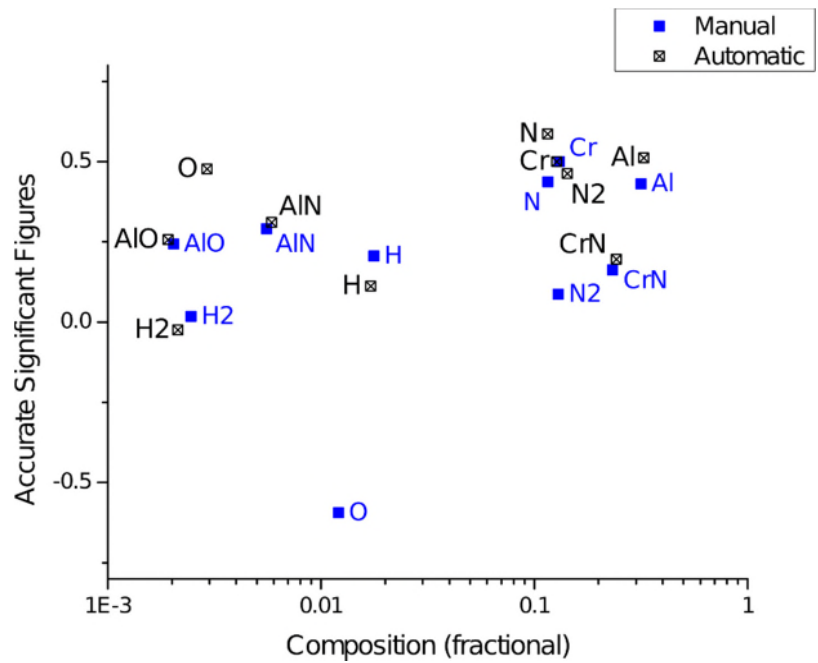

*Figure 7: Comparison of manual and assisted ranging for the Hard dataset. The most dramatic change is visible in the O signal, where in the manual case, the uncertainty is larger than the mean composition estimate.*

Figure 7 shows the results of the "Hard dataset". The dataset showed the lowest level of certainty from within the given trials. Participant agreement in this dataset was very poor, with uncertainty at sufficient levels to prevent accurate reporting of the leading digits for all species, for both the assisted and the manual attempts. The negative value for oxygen is due to the error being larger than the real value.

Timing for the assisted ranging did not show clear improvement across all tests, with the ratio of assisted:manual task completion time being estimated as 1.1, 0.60 and 0.7 for the easy, medium and hard datasets. It is noted that a standard user interface for ranging, not specifically suited for the tests undertaken here was used, which may further speed the assisted analysis.

# Discussion

## *Assisted Ranging*

The assisted ranging technique presented here shows that it is possible to strongly narrow the range of options available for each peak labelling. This method, consisting of a near-brute-force enumeration then a reduction stage, allows for a strongly narrowed set of possible peak labellings. A concern in both assisted and unassisted ranging is the bifurcation introduced in composition estimates when peaks are mislabelled.

Whilst it is currently possible to, from a practical perspective, reduce set sizes to usually 5-20 possible labellings (dependent upon material and isotopic complexity), this is insufficient for unique labelling.

Further work is required to reduce these set sizes, however the complexity of the physics, such as molecular stability is high. It is anticipated that a-priori data, such as from molecular dynamics simulations of small molecular clusters under high fields, may need to be included to improve the labellings further. To the authors' knowledge, there is no available literature for this. However, such or similar calculations have been done as in other vacuum science works [20]. An alternate route could be to utilise intensity and noise level information to check for side peak presence. This is however complicated by the possibility of overlaps.

## *Participant comparison*

It should be noted that in the discussion of "precision" in this work, only random error is accounted for. Systematic biasing mechanisms in atom probe – both in the field ion desorption process and within the atom probe apparatus itself [21][22][23] - are abundant and have been deliberately excluded from this work. The level of discrepancies in inter-operator comparisons for real datasets are quite large, with participant agreement for major components in the unassisted case rarely exceeding 95% confidence in the first 2 significant figures. The limited size of the dataset makes error estimation in the plots (i.e. the uncertainty on the uncertainty) difficult, hence in Figures 5, 6 and 7 no error bars are present.

For the "easy" dataset, the results show that the relative precision between participants (inter-participant) is significantly lower than for a single operator. Quantitatively, an individual operator achieves at most 2-3 significant figures for high-concentration species, whereas the inter-operator levels are between 1 and 2 significant figures. Automated ranging shows, for most species (excepting $C_2$, due to labelling bifurcation with auto-suggested FeO), a slightly higher level of inter-participant confidence than for manual ranging.

A key feature of the "medium" dataset is that the Ni and Fe accuracies are higher in the assisted case than for the manual case. The split originates from a dual-labelling of the Fe and Ni peaks, where different participants in the manual case ranged this drastically differently. These mis-labellings introduce large outliers in the dataset, which have a marked effect on the computed precision. As seen in the easy dataset, there appears to be a weak compositional dependence on precision. Somewhat disconcertingly, inter-operator precision for low-concentration species (at 95% confidence) dropped below 1 for species at less than 1% of the total dataset composition – i.e. the composition is essentially unknown. It is however acknowledged that the dataset size is smaller (Table 3) than for the other examples.

Finally the hard dataset, which shows a mixed result in terms of participant agreement for most species (excepting $N_2$ and O), shows very little overall agreement in both assisted and manual cases. For most species, participants did not agree. Disconcertingly, the scatter in concentration for O was found to be higher than the mean (hence the negative precision value).

It is furthermore likely that to reduce scatter in participants estimations, participants should undertake the tests in a more controlled environment. It is unlikely that the scatter can be reduced below that of an intra-operator test (Figure 4). It is believed however, that as participants become additionally skilled at analysis of the datasets, there would be some improvement in inter-participant precision.

# Summary

In this work, we have developed and implemented an algorithm for automated peak detection and identification in atom probe, by utilising existing packages for signal processing, and customised and robust algorithms for peak labelling. Such methods, currently not wide-spread in APT, are standard in other mass-spectroscopy techniques such as MALDI, owing to the higher complexity of large fragmented molecules.

Whilst in this work, we have also demonstrated a technique for achieving large reductions in the suggestion set size (*i.e.* possible labellings for a peak), further work is needed. At the simplest level, additional constraints in the suggested peaks (*e.g.* checking for smaller side-peaks presence, and relative background level) could be utilised with modest effort. There may be a significantly increased difficulty in developing further reductions, as new physical models for ion stability, due to the requirement that such models be simultaneously quantitative and highly generalisable. Analysis of peak intensities can be performed empirically, but is complicated by the difficulty of rigorously excluding all possible contaminant overlaps.

Furthermore, a study has been undertaken which examines the relative precision of participants who attempt to undertake ranging with and without these tools. Automated generation of rangefiles is shown to provide no loss of precision across groups of participants with and without the range data. Occasional improvements are seen, although there exists strong bifurcation of composition in cases of overlaps. It is believed that with further improvements in the predictive capacity of these algorithms, this result can be strongly improved.

As an outcome of this precision testing, it is estimated that inter-operator variance may be a large source of error, limiting reported APT compositions to between 0 and 2 significant figures, strongly falling off as a species composition decreases. One of the outstanding reasons for this can be mis-labelling of peaks, or only utilising one of several possible labels, the associated error for this can far outweigh variability due to range sizing.. This is a concerning result, which although highlighted previously by Hudson [3], emphasises the need for standardised interpretation of datasets. It is acknowledged that this is a recurring theme in APT literature, however it is believed that quantitative signal processing capabilities, such as those introduced here, are a possible solution. However, more work is required to remove the final modification step currently needed to be undertaken by operators, and to improve robustness in implementations at the peak detection stage. A pragmatic step may be to provide estimates for composition from different operators when working with new materials.

# Availability

The ranging programs presented here are available online, in source form at http://apttools.sourceforge.net/. The composition and background calculation is a "posgen" module, and the peak suggestion tool is undertaken using the "weights" program.

# Acknowledgements

The authors would like to thank the participants in this work for their time, as well as to those who